\documentclass[twocolumn,showpacs,preprintnumbers,amsmath,amssymb]{revtex4}
\usepackage{dcolumn}
\usepackage{bm}
\begin{document}
\draft
\def\bbox{\vrule height2mm depth0mm width5pt}
\def\theequation{\arabic{equation}}
\font\sqi=cmssq8
\def\DR{\rm I\kern-1.45pt\rm R}
\def\DC{\kern2pt {\hbox{\sqi I}}\kern-4.2pt\rm C}
 \pagenumbering{arabic}
\setcounter{page}{1}
\newcommand{\beq}{\begin{equation}}
\newcommand{\eeq}{\end{equation}}
\newcommand{\bea}{\begin{eqnarray}}
\newcommand{\eea}{\end{eqnarray}}
\newcommand{\bc}{\begin{center}}
\newcommand{\ec}{\end{center}}
\newcommand{\mb}{\mbox{\ }}
\newcommand{\bs}{\mbox{\boldmath $\sigma$}}
\newcommand{\bp}{\mbox{\boldmath $\pi$}}
\def\r{r^2_0}
\newcommand{\ch}{{\tt h}}
\newcommand{\ra}{\rightarrow}
\newcommand{\la}{\leftarrow}
\newcommand{\IR}{\mbox{I \hspace{-0.2cm}R}}
\newcommand{\IN}{\mbox{I \hspace{-0.2cm}N}}
\title{Quantum Mechanics Model on K\"ahler conifold}
\author{ Stefano Bellucci$^1$,    Armen Nersessian$^{2,3}$ and
Armen Yeranyan$^2$ }
\address{ $^1$ INFN-Laboratori Nazionali di Frascati,
 P.O. Box 13, I-00044, Italy\\
$^2$ Yerevan State University, Alex  Manoogian St., 1, Yerevan,
375025, Armenia\\
$^3$ Yerevan Physics Institute, Alikhanian Brothers St., 2, Yerevan, 375036,
 Armenia}

\begin{abstract}
\noindent
We propose an exactly-solvable model of the quantum oscillator on the class of
K\"ahler spaces (with conic singularities),
connected with two-dimensional complex projective
spaces. Its energy spectrum is nondegenerate in the orbital quantum
number, when the space has non-constant curvature.
We reduce the model to
a three-dimensional system interacting with the Dirac monopole.
Owing to noncommutativity of the reduction and quantization procedures, the Hamiltonian
of the reduced system gets non-trivial quantum corrections.
We transform the reduced system into a MIC-Kepler-like one
and find that quantum corrections arise only in its energy and
coupling constant. We present the exact spectrum of the
generalized MIC-Kepler system.
The one-(complex) dimensional analog of the suggested model
is formulated on the Riemann surface over the complex projective plane
and could be interpreted as a system with fractional spin.
\end{abstract}
\maketitle

\setcounter{equation}0
\noindent
{\bf Introduction.}
The quantum oscillator ranks as a most important
system of quantum mechanics, due to the existence of an
overcomplete set of hidden symmetries which  form a linear algebra.
The hidden symmetries provide the oscillator with unique properties,
e.g. a degenerate quantum-mechanical energy spectrum,
the separability of variables and exact solvability in a several coordinate
systems.
This allows one to
preserve the exact solvability even after some deformation of
the potential, or, at least, to
simplify the perturbative calculations.
These features make the oscillator to be a relevant system in a wide class
of problems in theoretical physics, including string/field theory,
gravity, condensed matter.
On the other hand, most problems in modern theoretical physics deal
with higher($d>3$)-dimensional curved spaces.
However, the quantum oscillator is generalized for spheres
and hyperboloids only \cite{higgs}, which have a constant curvature and no singularities.
For $d>2$ these spaces have no K\"ahler structure;
consequently, the corresponding oscillators have
a bad behavior with respect to supersymmetrization and
the inclusion of a constant magnetic field.
Moreover, in (super)string/brane and (super)gravity theories
K\"ahler spaces \cite{kahler}
and spaces with conic singularities (conifolds),
including K\"ahler conifolds, are of special importance
(see, e.g.,  \cite{cone} and refs therein).

In this note we propose a model of the four-dimensional quantum oscillator
on the $(\nu,\epsilon)$  parametric family of K\"ahler conifolds related with the
nonsingular cases of the complex projective space $\DC P^2$ (when $\nu =1$ and
$\epsilon=1$) and its noncompact version, i.e. the four-dimensional
Lobacewski space ${\cal L}_2$ (for $\nu =1$ and $\epsilon=-1$).
The K\"ahler structure is defined by the potential
\beq
K=\frac{r^2_0}{2\epsilon} \log (1+\epsilon(z\bar z)^\nu),\quad \nu
>0;
\quad \epsilon=\pm 1,
\label{kahler}\eeq
so that the  corresponding metric is
given by the expression
 \beq
 g_{a\bar b}=
\frac{\nu\r (z\bar z)^{\nu -1}}{2(1+\epsilon (z\bar z)^\nu )}
\left(
 \delta_{a\bar b}-\frac{1-\nu+
\epsilon (z\bar z)^\nu}{z\bar z\;
(1+\epsilon (z\bar z)^\nu)}\bar z^a z^b\right),
 \label{gkf}
\eeq
The scalar curvature takes the
 form
\beq
R=-\frac{4}{\nu\r}\frac{\nu -1 -\epsilon(2\nu+1)(z\bar z)^\nu}{z\bar z}\,.
\label{R}\eeq
We choose the following oscillator potential:
 \beq
V_{osc}=
\omega^2 g^{\bar a b}\partial_{\bar a}K \partial_b K=
\frac{\omega^2 \r}{2} (z\bar z)^\nu . \label{opf} \eeq
The exact classical
solvability of the model was established in \cite{cpn,ny}.
The potential (\ref{opf}) is distinguished
also with its respect to supersymmetrization and inclusion of a constant
 magnetic field \cite{cpn}.

The system is described by the Schr\"odinger equation
\begin{equation}
{\widehat{\cal H}}\Psi= E\Psi, \quad
{\widehat{\cal H}}= -\hbar^2 g^{a \bar b}\partial_a\partial_{\bar b} +V_{osc},
\label{o0}\end{equation}
where the metric and the potential are given by the expressions
 (\ref{gkf}), (\ref{opf}), respectively.
It is invariant under $U(2)$
rotations defined by the operators
 \beq
\begin{array}{c}
2\widehat{J}_0= z\partial -\bar
z\bar\partial ,\quad
2\widehat{\textbf{J}}= z\bs\partial-\bar \partial\bs\bar z,\cr
[J_0, {\bf J}]=0, \quad [J_i, J_k]=\imath\epsilon_{ikl}J_l\,,
\end{array}
\label{momvek}
\eeq
where $\bs$ are standard Pauli matrices, and $i,k,l=1,2,3$.
Here and further below we use the notation
$
 \partial_a ={\partial}/{\partial z^a}$,${\bar\partial}_{\bar a} =
{\partial}/{\partial {\bar z}^a}$.

We find that the system has remarkable properties.
It is exactly solvable: we find its {\it energy spectrum} and construct
{\it a complete
basis of wavefunctions}. On non-constant curvature spaces
the spectrum is nondegenerate in the orbital quantum number. Moreover, even on
constant curvature spaces the spectrum is nondegenerate in the eigenvalue
of the operator ${\widehat J}_0$.
Reducing the quantum Hamiltonian on the three-dimensional
 $(\nu,\epsilon)$-parametric space we find
that it gets a correction, with respect to the quantized three-dimensional Hamiltonian
reduced from four dimensions classically.
In other words, the reduction and quantization are noncommutative operations
in the proposed model.
The reduced system is specified by the presence of
a Dirac monopole field.
In the particular case $\nu=4$ it has no singularity, as its configuration space
is a three dimensional sphere/pseudosphere (two-sheet hyperboloid).
In this case the four-dimensional potential reduces to the  potential
of the oscillator on a (pseudo)sphere.
The reduced oscillator could be
converted into another exactly-solvable system
on the $(4\nu,-1)$-parametric three-dimensional space.
For $\nu=1$ it coincides
with the MIC-Kepler system (i.e. the
superintegrable generalization of the Coulomb system
with the Dirac monopole \cite{zwanziger})
 on the two-sheet hyperboloid \cite{np}.
Hence, the latter system  could be viewed as a
generalization of the MIC-Kepler system on nonconstant curvature conifolds.
Quantum corrections  change the only value of the coupling constant and
the energy, so that one could get the energy spectrum and the
wavefunctions of the MIC-Kepler like system from the oscillator ones!
The transformation to the Coulomb-like system does not have a mere academic interest.
Being related with the Hopf map, it has numerous applications in physics. The newest one is
the higher-dimensional quantum Hall effect \cite{hall}.
Notice, that the obtained results
could be straightforwardly extended to higher dimensions.\\

\noindent
{\bf Wavefunctions and spectrum.}The  Schr\"odinger equation
(\ref{o0}) could be separated
in the spherical coordinates
\beq
\begin{array}{c}
z^1=x^{1/\nu}\,
\cos{\frac{\beta}{2}}\,\exp{[\frac{\imath}{2}(\alpha+\gamma)]},
\cr
z^2=-\imath \, x^{1/\nu}\,
\sin{\frac{\beta}{2}}\,\exp{[-\frac{\imath}{2}(\alpha-\gamma)]},
\end{array}
\label{coordtrancp}
\eeq
upon the following choice of the wavefunction
\beq
\Psi=\psi(x) D^j_{m,s}(\alpha,\beta,\gamma)\label{sep}.\eeq
Here
$\alpha\in [0,2\pi)$, $\beta\in [0,\pi]$ and $\gamma\in
[0,4\pi)$, and   $x$ is a dimensionless radial coordinate
taking values in the interval  $ [0,\infty)$  for $\epsilon=+1$,
and in $[0, 1]$ for $\epsilon=-1$.
In the Wigner function $ D^j_{m,s}(\alpha,\beta,\gamma)$
$j$, $m$ denote orbital and azimuthal quantum numbers, respectively, while
$s$ is the eigenvalue of the operator $\widehat J_0$
\bea
&&\widehat{J}_0\Psi=s\Psi,\label{J0}\\
&&\widehat{\bf J}^2\Psi =j(j+1)\Psi,\quad
\widehat{J}_3\Psi=m\Psi ,\\
&& m,s=-j,-j+1,\ldots , j-1, j\;\;  j=0,1/2,1,\ldots \label{mj}
\eea
The  volume element reads
\beq
dV_{(4)}=\frac{\nu^2r_0^4}{32}\frac{x^3}{(1+x^2)^3}\sin\beta dxd\alpha
d\beta d\gamma. \label{vol}
\eeq
The   radial  Schr\"odinger equation looks as follows:
 \bea
 \frac{d^2\psi}{d
x^2}+\frac{3+\epsilon x^2}{1+\epsilon
x^2}\frac{1}{x}\frac{d\psi}{d x}
+\left[ \frac{2\r E+\epsilon{\omega^2 r_0^4} }{\hbar^2(1 +
\epsilon
x^2)^2}\right. -\nonumber\\
- \left. \frac{4\nu\,j(j+1)+
4(1-\nu)s^2}{\nu^2\,x^2(1+\epsilon
x^2)}-
\frac{\epsilon \delta^2}{1+\epsilon
x^2}
\right]\psi =0,
\label{rad2shf}
\end{eqnarray}
where
\beq
\delta^2=4\frac{s^2}{\nu^2}+\frac{\omega^2 r_0^4}{\hbar^2}.
\eeq
Making  the further substitution
 \beq
\begin{array}{cc}
 x=\tan\theta,&{\rm for}\; \epsilon=1\\
 x=\tanh\theta,&{\rm for}\; \epsilon=-1
\end{array},
\label{fgteta}\eeq
we shall  get the  regular wavefunctions, which form a complete
orthonormal basis,
\begin{widetext}
 \beq
\psi=\left\{
\begin{array}{c}
C\sin^{j_1-1}\theta\cos^{\delta}\theta\; _2F_1(-n,n+\delta+j_1+1;j_1+1;\sin^{2}\theta),\;{\rm for \;} \epsilon=1\;\;\\
C\sinh^{j_1-1}\theta\cosh^{-\delta+2n}\theta\; _2F_1(-n,-n+\delta,j_1+1,\tanh^{2}\theta),\;{\rm for \;} \epsilon=-1
\end{array}\right.\label{constant}\eeq
\end{widetext}
and the energy spectrum
 \bea
&&E_{n,\,j,\,s}
=\frac{\epsilon\hbar^2}{2\r}
\left(2n+ j_1+\epsilon\sqrt{\frac{4s^2}{\nu^2}+
\frac{\omega^2 r_0^4}{\hbar^2}}
+1 \right)^2-\nonumber\\
&&-\frac{2\epsilon\hbar^2}{\r}-\frac{\omega^2
\r}{2\epsilon}\,.
\label{Eo}\eea
Here
\beq
j_1^2=\frac{4j(j+1)}{\nu}+1 -\frac{4(\nu -1)s^2}{\nu^2}\,,
\label{j1}\eeq
whereas
\beq
n=\left\{\begin{array}{cc}0,1,\dots
,\infty&{\rm for}\;\epsilon=1\\
0,1,\dots
,n^{max}=[\delta/2-j-1]&{\rm for}\;\epsilon=-1
\end{array}\right.
\label{bound}\eeq
is the radial quantum number.
The normalization constants are defined by the expression
\begin{widetext}
\beq
\frac{\nu^2r^4_0\pi^2 n!\Gamma^2(j_1+1)}{4(2j+1)\Gamma(n+j_1+1)}
C^2=\left\{
\begin{array}{cc}
(2n+j_1+1+\delta)\Gamma(n+j_1+1+\delta)/\Gamma(n+1+\delta),
&{\rm for}\;\epsilon=1 ,\\
(\delta-2n-j_1-1)
\Gamma(\delta-n)/\Gamma (\delta-n-j_1),
&{\rm for}\;\epsilon=-1\;\; .
\end{array}\right.
\eeq
\end{widetext}
It is seen, that for $\nu\neq 1$ the energy spectrum is degenerate in the
azimuthal quantum number only.
The explicit dependence of
the spectrum on the  orbital quantum number $j$ is the
 quantum mechanical reflection
of the un-closedness of classical
 trajectories \cite{ny}.
On constant curvature spaces, where  $\nu=1$,
 the spectrum depends on  $s$ and $N\equiv 2n+2j$, i.e. it is degenerate in
the orbital quantum number $j$.
This degeneracy is due to the existence
of a hidden symmetry given by the operators
\beq
{\bf I}=\hbar^2 J\bs J^\dag +\bar z\bs z,\quad J_{a}=\imath\partial_a+\imath{\bar z}^a (\bar z\bar\partial ).
\eeq
 In the flat limit
($\r\rightarrow\infty$, $\theta\rightarrow0$ and $\r\theta=const$),
we get the correct formula for the four-dimensional oscillator energy spectrum
 \beq E_{N}=\hbar\omega (N+2),\quad N=2n+2j=
0,1,2,\ldots,\label{enerF}\eeq
i.e.  $N=2n+2j$ becomes the ``principal'' quantum number.
Notice that in the non-constant curvature case, $\nu\neq 1$ the
 infiniteness/finiteness
of the energy spectrum  is not straightforwardly correlated with the curvature,
as opposed to the case of
 constant curvature spaces (compare (\ref{R}) and (\ref{bound})).

The  two-dimensional counterpart of our model
 has a single complex coordinate $z$.
Performing the transformation $w=z^\nu$
we get the the Hamiltonian and the angular momentum
 operator on the Riemann surface over
$\DC P^1$ (for $\epsilon=1$) or Lobacewski
plane ${\cal L}$ (for $\epsilon=-1$)
\beq
\begin{array}{c}
{\cal H}=-\hbar^2(1+\epsilon w\bar w)^2 \partial_w\partial_{\bar w}
+\omega^2\r w\bar w,\cr
 2J= \nu (w\partial_w -\bar w\partial_{\bar w})\,,
\end{array}
\eeq
where  ${\rm arg} w\in [0,2\pi \nu )$.
   The energy of the system is
 given by the expression  (\ref{Eo}),
 where $s/\nu$ and $j_1$ are replaced by
 ${\tilde j}=j/\nu$.
For integer values of $\nu$
we could make a reduction by the $Z_\nu$ group and get a
 family
of  oscillators specified by the fractional
 spin $k=1/\nu, 2/\nu,\ldots, (1-1/\nu)$ (see \cite{ntt}).
It this case
${\tilde j}=k,k+ 1,     \ldots$ gets the meaning of the orbital quantum number
       on the
complex projective plane.
The arising of fractional spins
can be interpreted as a consequence of the presence of a magnetic flux tube
(see \cite{ntt}).
As opposed to the higher dimensional case, the spectrum is nondegenerate for any $\nu$,
which reflects the absence of hidden symmetry of the
system.\\

\noindent
{\bf Reduction and Coulomb-like systems.}
There is a well-known Kustaanheimo-Stiefel (KS) transformation  relating
the four-dimensional oscillator with the
three-dimensional Coulomb (and MIC-Kepler \cite{zwanziger}) system.
A similar transformation of the oscillator on
the four-dimensional sphere and (two-sheet) hyperboloid yields
 the MIC-Kepler system on the three-dimensional
hyperboloid \cite{np}. The quantum KS transformation
includes a reduction of the Schr\"odinger equation for the
four-dimensional oscillator by the $\widehat J_0$ operator, with
a subsequent transformation to the Schr\"odinger equation of the MIC-Kepler system.

In Ref.\cite{cpn} we applied a similar procedure to the classical
oscillator on $\DC P^2$ and found that,
as in the (pseudo)spherical case, it yields the
MIC-Kepler system on the three-dimensional hyperboloid.

Let us extend the KS transformation to the proposed model. We begin by
considering the reduction of the system to three dimensions.
For this purpose we consider the equation
(\ref{J0}) as a constraint and choose the functions below as
coordinates of the reduced system
\beq
\begin{array}{c}
{\bf x}=(z\bar z)^{\nu/2-1}z\bs\bar z,\quad [\widehat J_0, {\bf x}]=0,\cr
x_3=x\cos\beta,\;\quad x_2+ix_1=x\sin\beta{\rm e}^{i\alpha}.
\end{array}
\label{xr}\eeq
The wavefunction of the reduced system is related with the initial one as
follows:
\beq
\Psi_{(3)}(x,\alpha,\beta)=
\sqrt{x/(1+\epsilon x^2)}{\rm e}^{-is\gamma}\Psi.
\eeq
It is convenient to pass to the following coordinates:
\beq
{\bf y}=\left(\frac{\sqrt{1+\epsilon x^2}-1}{\epsilon x}\right)^{2/\nu}\frac{\bf x}{x}\Rightarrow x=\frac{2y^{{\sqrt
   \nu}/2}}{1-\epsilon y^{\sqrt\nu}},
\eeq
where the metric of the reduced space takes a conformally flat form
\beq
ds^2_{\nu, \epsilon, r_0}=
\frac{2\nu\r  y^{\sqrt{\nu}-2}(d{\bf y})^2}{(1+\epsilon y^{\sqrt{\nu}})^2}.
\label{mred}\eeq
Thus, we arrive at the reduced Schr\"odinger equation
\beq
{\widehat{\cal H}}_{\rm red}({\bf y},{\bf \pi})\Psi_{(3)}=E\Psi_{(3)},
\quad {\widehat{\cal H}}_{\rm red}=
{\widehat{\cal H}}^{0}_{red}+\hbar^2 {{\Lambda}}^{1},
\label{rs}\eeq
where
\bea
&&\widehat{\cal H}^{0}_{red}=
\frac{1}{\sqrt{g}}\widehat\pi_i{\sqrt{g}}g^{ij}\widehat\pi_j+
\hbar^2 s^2
\frac{(1+\epsilon y^{\sqrt{\nu}} )^4}{2\nu^2 r^2_0 y^{\sqrt{\nu}}
(1-\epsilon y^{\sqrt{\nu}})^2}+\nonumber\\
&&+
\frac{2\omega^2\r y^{\sqrt{\nu}}}{(1-\epsilon y^{\sqrt{\nu}})^2}
\eea
and
\beq
\Lambda^{1}=\frac{1}{8\r}
\left[\frac{4y^{\sqrt{\nu}}}{(1-\epsilon y^{\sqrt{\nu}})^2}-
\frac{3(1-\epsilon y^{\sqrt{\nu}})^2}{4y^{\sqrt{\nu}}}+10\epsilon\right].
\label{rocpn}\eeq
Here $\widehat{\bp}$ is the momentum
operator in the Dirac monopole field
\beq
\widehat{\bp}=-\imath\hbar\frac{\partial}{\partial {\bf y}}-
s{\bf A(y)},\quad [\widehat\pi_i,\widehat\pi_j]=
\hbar s\epsilon_{ijk}\frac{y^k}{y^3}.
\eeq
The energy spectrum is given by the same formula (\ref{Eo}) as in the four dimensional case,
with the only difference that $s=0,\pm 1/2,1,\ldots$
becomes a fixed parameter (i.e. the ``monopole number'').
Hence, instead of
(\ref{mj}) one  has
\beq
j=|s|,|s|+1,\ldots;\quad m=-j, -j+1,\ldots , j-1,j.
\eeq
Thus, we got a rather remarkable result:
{{\it
reducing the { quantum} Hamiltonian
yields a different outcome from { quantizing}
the reduced { classical} Hamiltonian.}
In the special case $\nu=4$ we get the system on the
three-dimensional sphere ($\epsilon=1$)
or pseudosphere (two-sheet hyperboloid), ( $\epsilon
=-1$).
In this case the reduced oscillator potential coincides with the
potential of the oscillator on the sphere (hyperboloid)
\cite{higgs}, so that in the absence of monopoles, $s=0$,
${\cal H}^0_{red}$ coincides with the Hamiltonian of the three-dimensional
Higgs oscillator.
Comparing the spectrum of the latter system \cite{KMP1} with that constructed above
(for $s=0$), one can see, that they coincide
only in the semiclassical limit. This is  due to the quantum correction $\hbar^2\Lambda_1$.

Now, we can complete the KS transformation,
dividing the both sides of the Schr\"odinger equation
(\ref{rs}) by $\r x^2$, and going to the
wavefunction ${\widetilde\Psi}_{(3)}=
 x^{-1/2}\Psi_{(3)}$.
 As a result, we shall get the  Schr\"odinger
equation of a MIC-Kepler-like  system on a
three-dimensional conifold with the metric
$ds^2_{{\nu}_1,-1, R_0}$ given by
(\ref{mred}), where  ${\nu_1}= 4\nu$, $R_0=\r$,
\beq
\widehat{\cal H}_{MIC}{\widetilde\Psi}_{(3)}={\cal
  E}\widetilde{\Psi}_{(3)},
\eeq
where the Hamiltonian $\widehat{\cal H}_{MIC}$ is of the form
\bea
\widehat{\cal H}_{MIC}=
\frac{1}{\sqrt{g}}\widehat\pi_i{\sqrt{g}}g^{ij}\widehat\pi_j+
2 s^2 \hbar^2
\frac{(1- y^{\sqrt{\nu_1}} )^2}{\nu^2_1 R^2_0 y^{
\sqrt{\nu_1}}}
-\nonumber\\
-
\frac{\gamma}{2 R_0}\frac{1+ y^{\sqrt{\nu_1}}}
{y^{\sqrt{\nu_1}/2}}\label{hmic},\eea
while  the energy
 and the coupling constant are given by the expressions
\beq
\gamma=\frac{E_{\rm osc}}{2}+\frac{\epsilon\hbar^2}{\r}(1-\frac{s^2}{\nu^2}),\quad
 -2{\cal E}=\omega^2+ \frac{\epsilon E_{osc}}{\r} +\frac{\hbar^2}{r^4_0}(1+2\frac{s^2}{\nu^2}).
\label{constants}\eeq
So, as opposed to the reduction procedure, where nontrivial quantum corrections
arise, the KS transformation of the four-dimensional
oscillator yields the MIC-Kepler like system on the
three-dimensional conifold, where the quantum corrections
have an impact on the coupling constant and the energy only.

Using the expressions (\ref{constant}),
we could convert the energy spectrum of the oscillator into that of the MIC-Kepler system
\beq
{\cal E}=-
\frac{2\left(\gamma
 -\epsilon\hbar^2(2n+j_1+1)^2/(4R_0)\right)^2}{\hbar^2(2n+j_1+1)^2}-
\frac{\epsilon\gamma}{R_0}
+\frac{\hbar^2}{2 R^2_0},
\eeq
where ${j}_1$ is defined by the expression (\ref{j1}).
{\it
We obtained an exactly solvable generalization of the MIC-Kepler
system on a class of three-dimensional spaces having
conic singularities and non-constant curvature.}
In the case of constant curvature, the system is degenerate
in the orbital quantum number $j$, otherwise it has an
explicit dependence on $j$.
\\

In our consideration we have used the common construction of the
quantum Hamiltonian on the curved configuration space, where the kinetic
energy term is replaced by the Laplace operator,
$$g^{ij}p_i\pi_j \to
\frac{1}{\sqrt{g}}\partial_i{\sqrt{g}}g^{ij}\partial_j,$$
which guarantees that  the Hamiltonian is Hermitean
and reparamerization-invariant.

In fact this definition assumes that the following realization of the
 momenta operators,
$\widehat\pi_i=-\hbar\partial_i$, which is non-Hermitean in the case of a
non-constant  metric. In Refs. \cite{mcm} the a priori
Hermitean realization of the momenta operators
 $\widehat\pi_i=-\hbar(\partial_i -\frac 14 \partial_i \log\det g)$ was suggested,
together with a slightly different definition of the observables quadratic on momenta
(and, consequently, of the Hamiltonian) respecting the reduction procedure.
Upon this definition of momenta, both the initial and the reduced
quantum Hamiltonian will get quantum corrections, with respect to the Hamiltonian
as we defined it in our conventions.
It seems to be interesting to compare the spectra and the properties
of the systems under consideration in both approaches.
(We would like to thank the Referee for pointing out
Refs. \cite{mcm} and drawing our attention to
such questions).
\\

\noindent
{\bf Summary and Conclusion.} Let us summarize our results.
We proposed an exactly-solvable model of a quantum
oscillator on two-(complex) dimensional complex projective space
(in its compact and noncompact versions), as well as on the
related non-constant curvature K\"ahler spaces with conic singularities,
parameterized by $\nu>0$ and $\epsilon=\pm 1$.
We reduced the oscillators to those on $(\nu,\epsilon)-$
parametric family of three-dimensional non-constant curvature conifolds
related with the three dimensional (pseudo)sphere.
The reduced systems are specified by the presence of a Dirac monopole.
%
During the reduction the quantum Hamiltonian gets additional corrections
with respect to the quantized Hamiltonian, reduced from four-dimensional space at
the classical level.
Then, we 
 transformed the  reduced
$(\nu,\epsilon)-$ oscillator in the  MIC-Kepler like system
on  $(4\nu,-1)-$ conifold and get its exact energy spectrum.
Opposite to the reduced oscillator,
in the final system  quantum corrections affect the
 energy and coupling constant only. 
The two-dimensional counterpart of the suggested model corresponds to a system
with fractional spin
on the  complex projective plane. Its spectrum is non-degenerate for any $\nu$.

It appears that the proposed system preserves its
exact solvability after inclusion of a {\it constant magnetic field}.
It is also a distinguished system with respect to supersymmetrization, as
it admits a {\it non-standard} ${\cal N}=4$ supesymmetric extension,
 which respects the inclusion of ca onstant magnetic field (cf.
 \cite{cpn}).
We are planning to discuss such matters in detail elsewhere.\\

\noindent
{\bf{ Acknowledgments}}
We are indebted to  R. Avagyan and
L. Mardoyan for useful
conversations. Special thanks to L. Mardoyan for help in calculating
the normalization constants (\ref{constant}).
The work of S.B. was supported in part
by the European Community's Human Potential
Programme under
contract HPRN-CT-2000-00131 Quantum Spacetime,
the INTAS-00-00254 grant and the
NATO Collaborative Linkage Grant PST.CLG.979389.
The work of A.N. was supported by grants INTAS 00-00262  and ANSEF  PS81.

\end{document}